\documentclass[aps,pre,twocolumn,letterpaper,floatfix,showpacs]{revtex4-1}
\usepackage{graphicx} 
\usepackage{amsmath,amssymb,amsfonts} 
\usepackage{mathtools}
\usepackage{epstopdf}
\usepackage{natbib}

\begin{document}

\title{A minimal model for slow, sub-Rayleigh, super-shear and unsteady rupture propagation along homogeneously loaded frictional interfaces}

\author{Kjetil Th\o gersen$^\text{a,b}$}
\email{kjetil.thogersen@fys.uio.no}
\author{Henrik Andersen Sveinsson$^\text{a,c}$}%
\author{David Sk{\aa}lid Amundsen$^\text{d}$}
\author{Julien Scheibert$^\text{e}$}
\author{Fran\c{c}ois Renard$^\text{a,b,f}$}
\author{Anders Malthe-S{\o}renssen$^\text{a,c}$}
\affiliation{
Physics of Geological Processes, The NJORD Centre, University of Oslo, 0316 Oslo, Norway$^\text{a}$\\
Department of Geosciences, University of Oslo, 0316 Oslo, Norway$^\text{b}$\\
Department of Physics, University of Oslo, 0316 Oslo, Norway$^\text{c}$ \\
Astrophysics Group, University of Exeter, EX4 4QL Exeter, UK$^\text{d}$\\
Univ Lyon, Ecole Centrale de Lyon, ENISE, ENTPE, CNRS, Laboratoire de Tribologie et Dynamique des Syst{\`e}mes LTDS, F-69134, Ecully, France$^\text{e}$\\
University Grenoble Alpes, University Savoie Mont Blanc, CNRS, IRD, IFSTTAR, ISTerre, 38000 Grenoble, France$^\text{f}$}
\date{\today}

\begin{abstract}
In nature and experiments, a large variety of rupture speeds and front modes along frictional interfaces are observed. Here, we introduce a minimal model for the rupture of homogeneously loaded interfaces with velocity strengthening dynamic friction, containing only two dimensionless parameters; $\bar \tau$ which governs the prestress, and $\bar \alpha$ which is set by the dynamic viscosity. This model contains a large variety of front types, including slow fronts, sub-Rayleigh fronts, super-shear fronts, slip pulses, cracks, arresting fronts and fronts that alternate between arresting and propagating phases. Our results indicate that this wide range of front types is an inherent property of frictional systems with velocity strengthening branches.
\end{abstract}

\maketitle

\section{Introduction}

The onset of sliding of frictional contacts is often mediated by the propagation of a slip front along its interface, in natural, laboratory and industrial situations \cite{scholz1998earthquakes,vakis2018modeling,svetlisky2019brittle}. Slip fronts typically nucleate at the weakest and/or most loaded part of the interface, propagate and eventually either invade the whole contact or arrest after breaking only a portion of the interface.

This front propagation can be characterized by two main features: front speed and front mode. Two main front modes have been identified, both in earthquakes and in laboratory experiments: Cracks where the interface behind the front slips until propagation ends \cite{rubinstein2004detachment,xia2005laboratory,ben2010dynamics,latour2011ultrafast}, and slip pulses where the ruptured part of the interface rapidly heals and re-sticks during propagation  \cite{baumberger2002self,lykotrafitis2006self,nielsen2010experimental,shlomai2016structure,latour2011ultrafast}. Propagation can occur at speeds differing by orders of magnitude; at velocities close to but below the Rayleigh wave speed (sub-Rayleigh), above the shear wave speed (super-shear), and at speeds orders of magnitude smaller than the sound speeds (slow).  In addition, quasi-static fronts with a speed controlled by the external loading rate \cite{prevost2013probing} have been reported.  For dynamic cracks (from slow to super-shear, through sub-Rayleigh), higher propagation speeds are found for larger prestress of the interface \cite{ben2010dynamics} and for larger dynamic stress drop \cite{svetlizky2017brittle}. Such observations are not limited to experiments. In nature, earthquakes can propagate at both seismic and aseismic velocities \cite{burgmann2018geophysics}, which obey different relations between seismic moment and earthquake duration \cite{ide2007scaling}. Observations in nature also include periodic pulsing of aseismic events \cite{nadeau2004periodic}. 

When the propagation speed decreases to zero before a front reaches the edge of an interface, the front is denoted as arrested. Such fronts can be considered as precursors to sliding if they precede fronts spanning a larger portion of the interface \cite{rubinstein2007dynamics}. The propagation length, like the propagation speed, depends on both the interfacial prestress and dynamic stress drop \cite{bayart2016fracture}. Overall, the combination of the front mode, the range of its propagation speed and the information about whether it has arrested constitutes what we call the front type.

The range of observed front types have already been successfully reproduced by a variety of models. Arrested cracks have been reproduced using quasi-static models \cite{scheibert2010role,braun2014propagation,kammer2015linear,bayart2016fracture}, or elastodynamic models in 1D \cite{maegawa2010precursors,amundsen20121d} or 2D \cite{tromborg2011transition,radiguet2013survival,tromborg2014slow,taloni2015scalar,tromborg2015speed}, assuming either continuous \cite{scheibert2010role,maegawa2010precursors,tromborg2011transition,amundsen20121d,radiguet2013survival,taloni2015scalar} or discrete-microcontact-based friction laws \cite{tromborg2014slow,tromborg2015speed}, or fracture concepts \cite{kammer2015linear,bayart2016fracture}. Slip pulses have been reproduced using discrete \cite{gerde2001friction} or continuum models assuming either a Coulomb~\cite{yastrebov2016sliding}, regularized Coulomb \cite{andrews1997wrinkle,cochard2000fault,adda2003self} or state-and-rate \cite{putelat2017phase,brener2018unstable} friction laws. Models of cracks are ubiquitous, featuring super-shear \cite{tromborg2011transition,amundsen2015steady,kammer2018equation}, sub-Rayleigh \cite{tromborg2011transition,radiguet2013survival,bar2013instabilities,tromborg2014slow,tromborg2015speed}, slow~\cite{bouchbinder2011slow,bar2012slow,tromborg2014slow,tromborg2015speed} or quasi-static \cite{braun2009dynamics,tromborg2014slow,papangelo2015fracture} fronts. Note that front speed has been shown to depend on many features of the frictional system, including slip history \cite{radiguet2013survival,tromborg2015speed}, interaction between different fault planes \cite{romanet2018fast}, the shape of the high speed branch of the friction law \cite{bar2015velocity}, and spatial heterogeneities in stress or constitutive parameters \cite{boatwright1996frictional,bizzarri2001solving,liu2005aseismic,bruhat2016rupture,helmstetter2009afterslip}.

In front of so many different models, the physical origin of the observed richness in front types remains elusive. In this paper we address the question of the single minimal model that would reproduce the observed richness. We show how a simple friction model, reducible to only two non-dimensional parameters, contains a wide range of observed front types. Our findings indicate that the richness of front types is an inherent property of interfaces with velocity strengthening dynamic friction, which is a generic feature of both dry \cite{bar2014velocity} and lubricated interfaces \cite{gelinck2000calculation,olsson1998friction}.

\begin{figure}
\centering
\includegraphics[width = .5\textwidth]{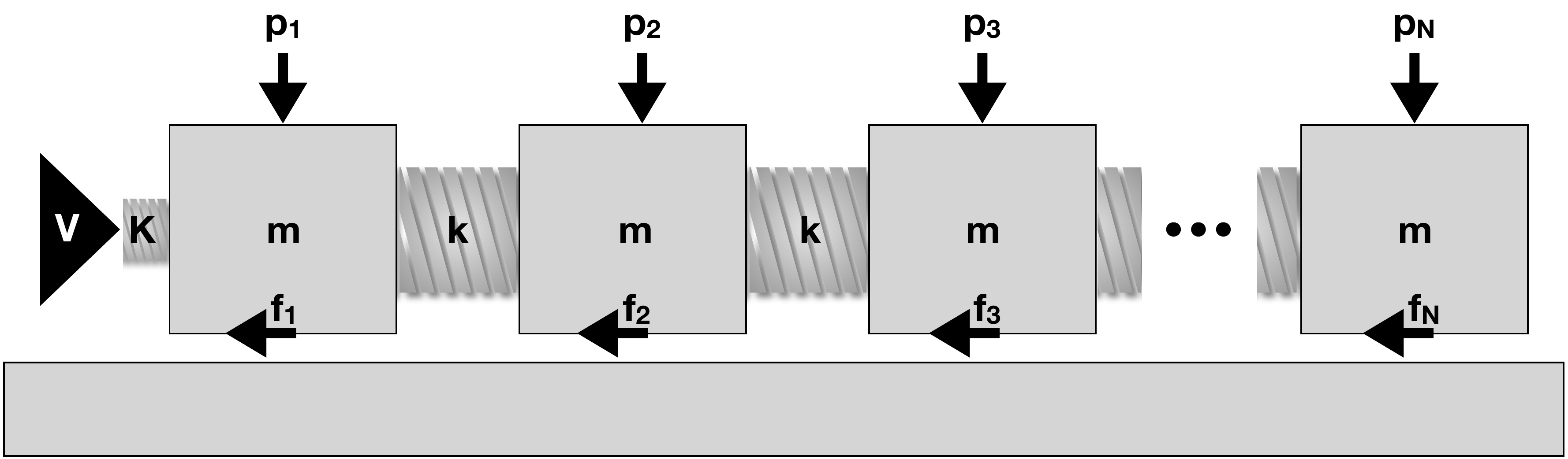}
\caption{System sketch. We solve the Burridge-Knopoff model in the limit of soft tangential loading (small $K$ and $v$), for a prestressed interface with Amontons-Coulomb friction with velocity strengthening dynamic friction. $V$ is the driving velocity, $K$ is the driving spring constant, $m$ is the block mass, $p_i$ is the normal force on block $i$, and $f_i$ is the friction force on block $i$. Assuming homogeneous stress at the interface and soft tangential loading, this results in a system with two dimensionless parameters as described by equation \ref{eq:EoM}.
\label{fig:sketch}}
\end{figure}
\begin{figure*}
\centering
\includegraphics{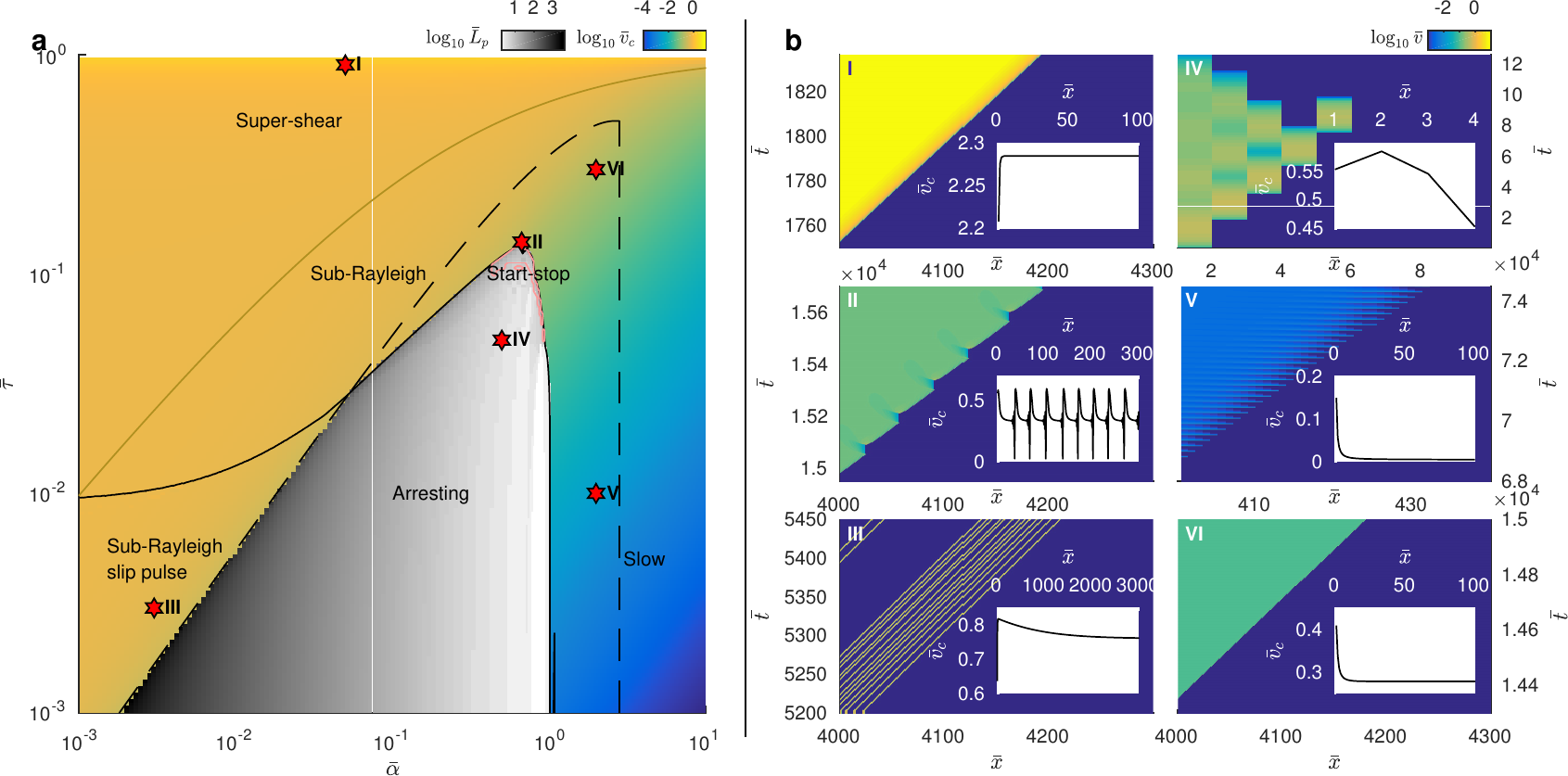}
\caption{a: Phase diagram in $\bar \alpha$ and $\bar \tau$ of the steady state front velocities. The dashed black line shows the line of unconditional propagation given by equation \ref{eq:uncond_prop} (derived in appendix \ref{sec:app_tau_uncond}). The solid black line shows the region where blocks are found to arrest, which is obtained numerically. The limit of super-shear velocities is marked with a solid yellow line. The region where start-stop fronts are found is marked with a pink contour. The grayscale colormap shows the propagation length $L_p$ of arresting fronts. Right: examples of super-shear cracks (I), unsteady rupture (II), slip pulse propagation (III), arresting fronts (IV), slow cracks (V) and sub-Rayleigh cracks (VI).
\label{fig:phasediagram}}
\end{figure*}
\begin{figure}
\centering
\includegraphics{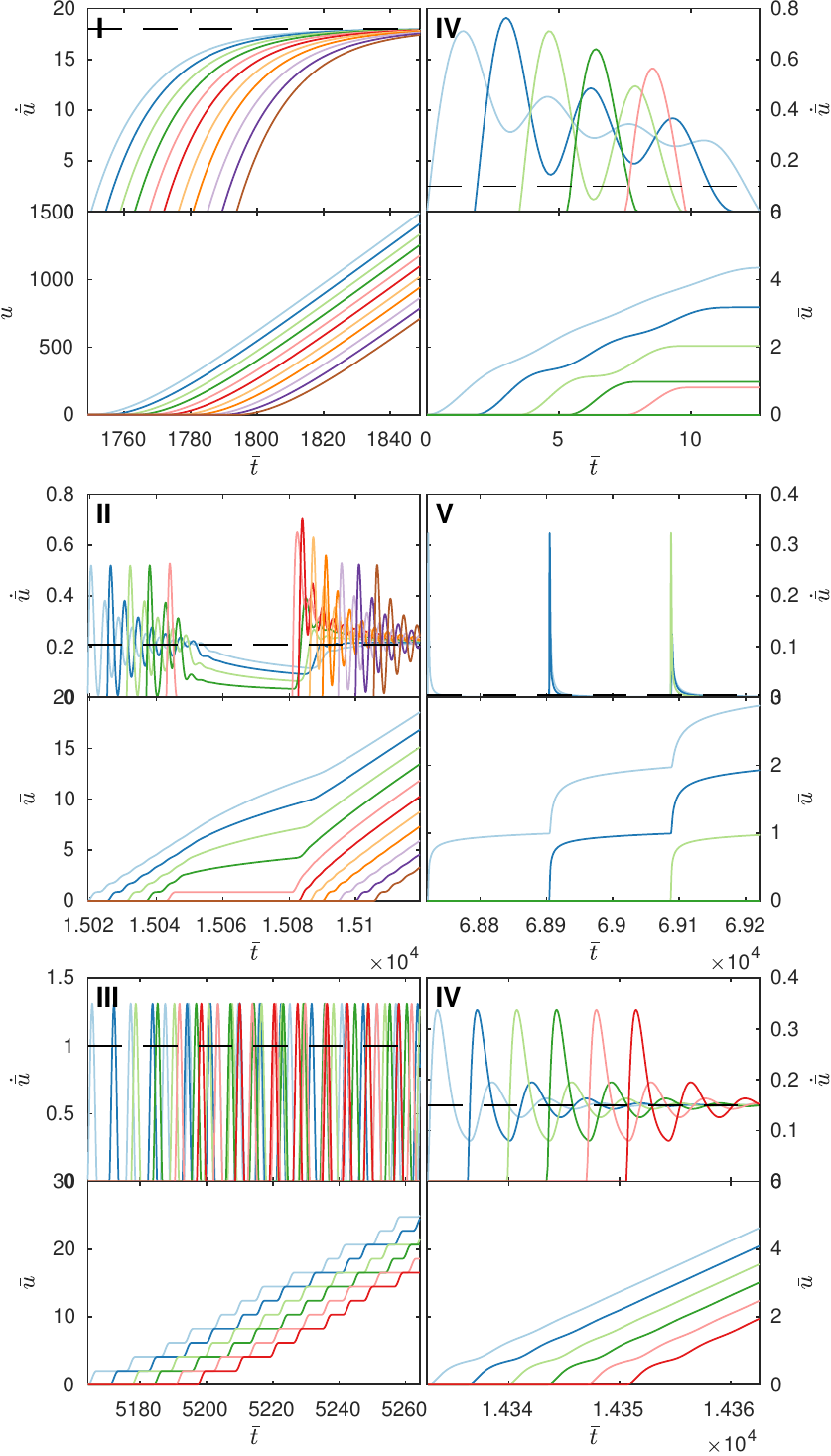}
\caption{Local slip and slip speed for simulations corresponding to \figurename~\ref{fig:phasediagram} The color coding (from blue to brown, through green, red and orange) of corresponds to increasing block index, and the lines are separated by (I) 10 blocks, (II) 2 blocks, (III) 5 blocks, (IV) 1 block, (V) 1 block and (VI) 1 block. Black dashed lines show the steady state slip velocity $\bar \tau / \bar \alpha$.
\label{fig:localSlip}}
\end{figure}

\section{Model description}
We introduce a one-dimensional Burridge-Knopoff model \cite{burridge1967model} for homogeneously loaded interfaces obeying Amontons-Coulomb friction, where the dynamic friction coefficient is velocity strengthening (\figurename~\ref{fig:sketch}). The dimensionless equation of motion for sliding blocks is derived in appendix \ref{sec:app_eom} and can be written as
\begin{align}
    \ddot{\bar{u}}_i -\bar{u}_{i-1} - \bar{u}_{i+1} + 2\bar{u}_i + \bar{\alpha} \dot{\bar{u}}_i - \bar\tau = 0, \quad \forall i \in [1,N],
    \label{eq:EoM}
\end{align}
where $\bar u$ is the dimensionless displacement. $\bar \tau =  \frac{\tau/p - \mu_\text{k}}{\mu_\text{s} - \mu_\text{k}}$ is the dimensionless prestress, where $\tau$ is the shear preload, $p$ is the normal load and $\mu_\text{k}$ and $\mu_\text{s}$ are the dynamic and static friction coefficients, respectively. $\bar \alpha = \frac{\alpha}{\sqrt{km}},$ is the dimensionless viscosity, where $\alpha$ is the viscosity coefficient of the interface, $k$ is the spring constant between two blocks, and $m$ is the block mass. We select the dimensionless time $\bar t$ and the dimensionless block separation so that the dimensionless speed of sound in the model is $\bar v_s = 1$ . We assume soft tangential loading (small driving velocity $V$ and small driving spring stiffness $K$), which results in boundary conditions given by a constant force on the leftmost block equal to its value when that block reaches its static friction threshold; $\bar{u}_{0} = \bar u_1 + 1 - \bar \tau$. At the right boundary we keep block $N+1$ fixed; $\bar{u}_{N+1} = 0$ (appendix \ref{sec:app_eom}).

Blocks start to move once the static friction threshold is reached, which in dimensionless units can be written as
\begin{align}
 \bar{u}_{i-1} + \bar{u}_{i+1} - 2\bar{u}_i \geq 1 - \bar \tau
 \label{eq:static_threshold}
\end{align}
Blocks restick if the velocity $\dot{\bar u}$ changes sign.

The assumed friction law has a discontinuity at $v=0$, because $\mu_s \neq \mu_k$. Note that we investigated regularization of the model using either a characteristic length scale or a characteristic velocity scale (appendix~\ref{sec:app_regularization}). The overall qualitative features of the model, in particular the various front types produced and their occurrence as a function of $\bar \alpha$ and $\bar \tau$, are the same as in the simple, unregularized model (\figurename~\ref{fig:phasediagram_regularized}). At large slip velocities, we assume a velocity strengthening friction force, as is classical for both lubricated \cite{diew2015stribeck} and dry interfaces~\cite{bar2014velocity}. The combination of a velocity weakening branch followed by a velocity strengthening branch as the slip velocity in increased is typical for Stribeck-like curves \cite{gelinck2000calculation,olsson1998friction}.

We emphasize that the present model can be fully described using only two dimensionless numbers: the dimensionless viscosity, $\bar \alpha$, which defines the velocity strengthening term and the prestress $\bar \tau$, which indicates how close the interface is to its static friction threshold.

\section{Richness of slip and rupture}
We have performed  $4 \times 10^4$ simulations for  $\bar \tau \in [10^{-3},1)$ and  $\bar \alpha \in [10^{-3},10]$ to obtain the relationship between prestress, viscosity and front velocities shown in \figurename~\ref{fig:phasediagram}. To reduce the computational cost we have performed $2 \times 10^4$ for $\bar \alpha \in [10^{-3},0.5]$, with $N = 5000$ and $2 \times 10^4$ for $\bar \alpha \in [0.5,10]$ with $N = 100$ (the transients in the large $\bar \alpha$ regime require a smaller propagation distance before steady state is reached). The simulations were run until all blocks stopped or the front reached the end.

\subsection{Front speed}
\begin{figure}
\centering
\includegraphics{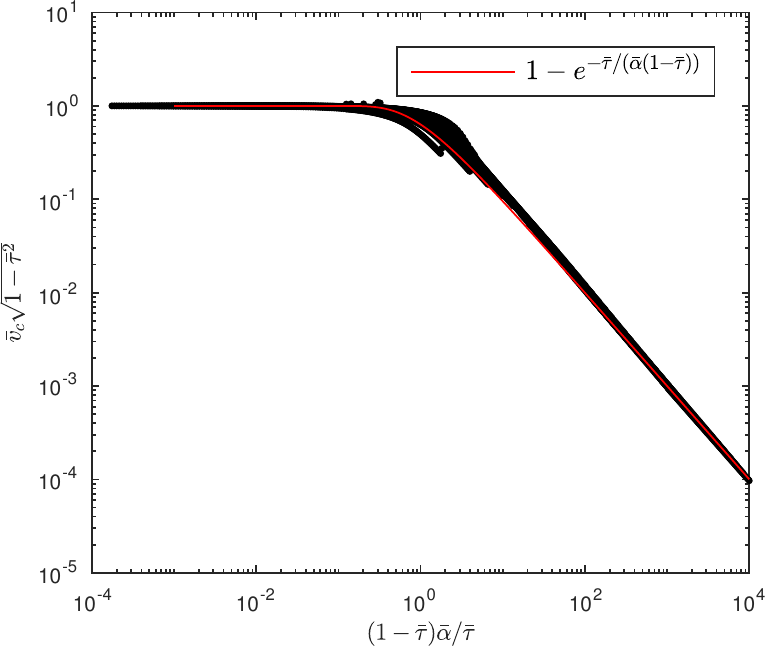}
\caption{Rescaled steady state propagation speed using equations \ref{eq:vc_alpha=0} and \ref{eq:vc_alpha>>0}. From the collapse we can determine an approximate solution of the steady state velocity for all $\bar \tau$ and $\bar \alpha$ which is given by equation \ref{eq:vc_approx} and plotted as a red line in the figure.
\label{fig:velocityCollapse}}
\end{figure}

For each simulation, we have measured the steady state velocity for fronts progagating through the entire interface (colorscale in \figurename~\ref{fig:phasediagram}a). To obtain steady state front velocities, we measure the times of rupture of all blocks, and extrapolate $\bar v_{c}(1/\bar x)$ linearly to $\frac{1}{\bar x} = 0$ using the last 20\% of the blocks. For arresting fronts, we measure the propagation length $\bar L_p$. The results are shown in \figurename~\ref{fig:phasediagram}, with corresponding slip and velocity curves for the examples shown in \figurename~\ref{fig:localSlip}. The front velocities span a continuum from slow velocities for low $\bar \tau$ and large $\bar \alpha$ to super shear-velocities at large $\bar \tau$ and low $\bar \alpha$. The front velocity at $\bar \alpha = 0$  can be found analytically and is given by  \cite{amundsen2015steady}
\begin{equation}
\bar v_c(\bar \alpha = 0) = \frac{1}{\sqrt{1-\bar \tau^2}}
\label{eq:vc_alpha=0}
\end{equation}
To estimate the steady state propagation speed in the limit of large $\bar \alpha$, we start with the steady state slip speed, which can be obtained directly from equation \ref{eq:EoM}. If the slip speed is constant, then $\ddot {\bar{u}}_i = 0$, and $\bar u_{i-1}-\bar u_{i+1}+2\bar u_i \approx 0$, so that equation \ref{eq:EoM} reduces to $\bar \alpha \dot {\bar u}_\text{ss} - \bar \tau = 0$, where the steady state slip speed is
\begin{equation}
\dot {\bar u}_\text{ss} =  \bar \tau / \bar \alpha.
\end{equation}
In the limit of large $\bar \alpha$ we expect the propagation speed to be governed by the steady state slip speed. 
If block $i$ has just ruptured, the displacement necessary to trigger the rupture of block $i+1$ is $1 - \bar \tau$ ($\bar u_{i+1}=0$ and $\bar u_{i} = 0$ in equation \ref{eq:static_threshold}). The dimensionless distance between the blocks is $1$. At a speed of $\dot{\bar u}_\text{ss} = \frac{\bar \tau}{\bar \alpha}$  it takes a dimensionless time $\frac{\bar \alpha}{\bar \tau}(1-\bar \tau)$ to travel the dimensionless distance $1-\bar \tau$, so that the front speed is
\begin{equation}
\bar v_c(\bar \alpha \gg 0, \bar \tau \ll 1) \approx \frac{\bar \tau}{\bar \alpha(1-\bar \tau)}.
\label{eq:vc_alpha>>0}
\end{equation}
In \figurename~\ref{fig:velocityCollapse} we use equations \ref{eq:vc_alpha=0} and \ref{eq:vc_alpha>>0} and find that we obtain a decent data collapse of the steady state front velocities when we plot $\bar v_c(\bar \alpha = 0)$ against $\bar v_c(\bar \alpha \gg 0, \bar \tau \ll 1)$. From this collapse we obtain an empirical approximation of the front propagation velocity which is valid in both limits
\begin{equation}
\bar v_c \approx \frac{1 - e^{- \frac{\bar \tau}{\bar \alpha (1- \bar \tau)} }}{\sqrt{1- \bar \tau^2}}.
\label{eq:vc_approx}
\end{equation}
Inserting for the dimensional quantities in equations \ref{eq:vc_alpha=0} and \ref{eq:vc_alpha>>0}, we find the following dependencies on the density $\rho$:
\begin{align}
v_c(\bar \alpha = 0) \propto \frac{1}{\sqrt{\rho}} \\
v_c(\bar \alpha \gg 0, \bar \tau \ll 1) = \text{constant}
\label{eq:frontVel_rho}
\end{align}
From this we can immediately conclude that fast fronts are dominated by inertia while slow fronts are not. We emphasize that this separation between inertial and non-inertial fronts only apples to the end-member solutions of equation \ref{eq:vc_approx}, and does not apply for the intermediate front velocities (found for large $\bar \tau$ and $\bar \alpha$ in \figurename~\ref{fig:phasediagram}).

\subsection{Front type}
We observe that the model is able to produce a large variety of front types. In addition to sub-Rayleigh, super-shear, and slow cracks, we observe slip pulse solutions, arresting fronts, as well as rupture speeds that alternate between propagating and arresting phases. \figurename~\ref{fig:phasediagram} contains boundaries of the different front types, as well as examples of space-time development of each front type. The corresponding local slip and slip speed for the examples in \figurename~\ref{fig:phasediagram} is shown \figurename~\ref{fig:localSlip}. 

The sub-Rayleigh and super-shear front velocities are found in the limit of small $\bar \alpha$, or large $\bar \alpha$ combined with large $\bar \tau$. The front velocities systematically increase with increasing $\bar \tau$. Examples of sub-Rayleigh and super-shear propagation is found in \figurename~\ref{fig:phasediagram}bI and \figurename~\ref{fig:phasediagram}bVI, with corresponding slip profiles found in \figurename~\ref{fig:localSlip}I and \figurename~\ref{fig:localSlip}VI. The slow fronts are found in the limit of large $\bar \alpha$ and small $\bar \tau$. An example of a slow front is shown in \figurename~\ref{fig:phasediagram}bV with the corresponding slip profiles found in \figurename~\ref{fig:localSlip}V.

We also observe a large region in $(\bar \tau, \bar \alpha)$ where steady-state solutions do not exist (greyscale in \figurename~\ref{fig:phasediagram}). For these arresting fronts we measured the propagation distance $\bar L_p$, which increases with decreasing $\bar \alpha$. There is also a sharp transition from fronts that stop within a small distance and the slow regime where steady state-solutions exist close to $\bar \alpha \simeq 1$.

Slip pulse solutions in the Burridge-Knopoff model typically manifest as a series of slip bands, each a few blocks wide, propagating at the same velocity. The steady state slip pulse-region is found for small $\bar \alpha$ and small $\bar \tau$, but the arresting region also contains slip pulse solutions. An example of a slip pulse is shown in  \figurename~\ref{fig:phasediagram}bIII, with the corresponding slip profiles in \figurename~\ref{fig:localSlip}III.

We also observe front propagation that alternates between propagating and arresting phases, which we denote as start-stop fronts. The mechanism behind this front type is as follows: If a crack that is arresting is sufficiently long, it will always be able to restart as long as all blocks behind the front are still sliding.
If a block at the front of a propagating crack arrests at position $\bar u = (1-\bar \tau)-\bar \epsilon$, the block in front of it will carry a stress of $1-\bar \epsilon$, where $1$ corresponds to the static friction threshold. There is thus a possibility for a force $2-\bar \epsilon$ to be carried by two arrested blocks at the front tip. Restarting the propagation requires that there is sufficient force available in the form of slow slip behind the front tip. The available force can be written as  $\bar l \bar \tau + (1-\bar \tau)$ where $\bar l$ is the position of the front tip at the time of arrest. The criterion for the existence of a start-stop front can be found by balancing these two contributions; $\bar l \bar \tau + (1-\bar \tau) \geq 2-\bar \epsilon$. The criterion for the unconditional restart of a crack that has arrested is found when $\bar \epsilon \rightarrow 0$, which corresponds to a stress close to the static friction threshold on the two arrested blocks in front of the crack. This can be written as a crack length $\bar l$ that allows for the existence of start-stop fronts
\begin{equation}
\bar l \geq \frac{1}{\bar \tau} + 1.
\label{eq:startstop}
\end{equation}
Note that this argument requires that the entire interface behind the front tip is sliding, which means that slip pulses will not be subject to this behavior. 
The start stop fronts are marked with a pink contour in \figurename~\ref{fig:phasediagram}a and an example is shown in \figurename~\ref{fig:phasediagram}bII with the corresponding slip profiles found in \figurename~\ref{fig:localSlip}II, and the measured $\bar L_p$ is taken as the propagation length when the front stops for the first time ($\bar L_p = \bar l$).

\subsection{Phase diagram boundaries}
In the following, we investigate the boundaries between the different front types observed in \figurename~\ref{fig:phasediagram}a. 

First, we find the line of unconditional propagation in \figurename~\ref{fig:phasediagram} (dashed), which separates the slip pulse region from the arresting region at small $\bar \alpha$ and then divides the sub-Rayleigh region for larger $\bar \alpha$. If a block at the front tip is able to trigger the next block even though the block behind it has stopped, a propagating front will not be able to arrest. Solving for this criterion in $\bar \tau$ and $\bar \alpha$ gives a criterion $\bar \tau_\text{uncond} (\bar \alpha)$ above which steady state propagation will always occur. The condition of the existence of such solution can be found analytically (appendix \ref{sec:app_tau_uncond}) and is given as
\begin{align}
	\bar \tau_\text{uncond} = \frac{1}{2} -\frac{1}{2}e^{-\frac{\pi \bar \alpha}{\sqrt{8 - \bar \alpha^2}}}.
        \label{eq:uncond_prop}
\end{align}

\begin{figure}
\centering
\includegraphics{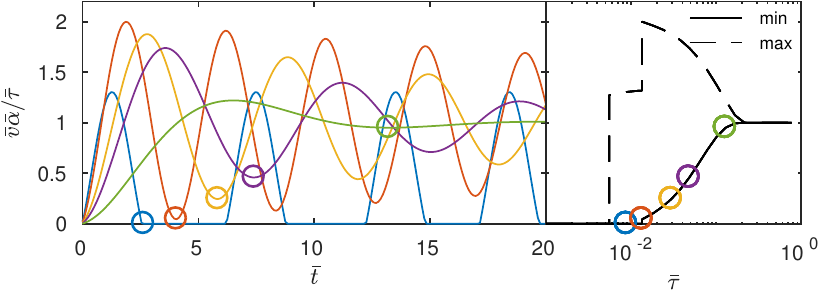}
\caption{Left: Velocity oscillations for block 4000 as a function of time with varying $\bar \tau$ and $\bar \alpha = 0.01$. Right: Amplitude of the oscillations as a function of $\bar \tau$ with colors corresponding to the data shown in the left panel. The transition from continuous sliding behind the front to slip pulses is determined by the oscillation amplitude compared to the steady state slip speed $\bar \tau / \bar \alpha$. Arresting fronts are found if steady state slip pulses cannot occur.
\label{fig:oscillationAmplitude}}
\end{figure}

To obtain the arresting domain of the phase diagram, we need to determine when blocks in the system are able to reach zero velocity. This can occur during very short transients or because a steady state solution contains large velocity fluctuations.  We have not been able to determine this criterion analytically, but it is straightforward to find the criterion numerically.  Blocks can either stop at the front tip as in \figurename~\ref{fig:localSlip}IV, or because of velocity oscillations behind the front, as demonstrated in \figurename~\ref{fig:oscillationAmplitude}. For a fixed $\bar \alpha$, varying $\bar \tau$ systematically changes the amplitude of such oscillations, which leads to a well defined criterion for the existence of arresting blocks $\bar \tau_\text{arrest}(\bar \alpha)$. The procedure for determining the criterion is as follows: We use a system of $100$ blocks. For a given $\bar \alpha$ and $\bar \tau$, we run a simulation and check whether it contains blocks that start and then arrest before the front reaches the end. We then use the bisection method for fixed $\bar \alpha$, varying $\bar \tau$ to find the limiting $\bar \tau_\text{arrest}$.
This solution $\bar \tau_\text{arrest}(\bar \alpha)$ is plotted as a solid black line in \figurename~\ref{fig:phasediagram}.  

The two criteria $\bar \tau_\text{uncond}$ and $\bar \tau_\text{arrest}$ combined explains both the region of the phase diagram where steady state slip pulse solutions exist and the location of the arresting region. Steady state slip pulses exist in for $\bar \tau (\bar \alpha) \in [\bar \tau_\text{uncond}(\bar \alpha),\bar \tau_\text{arrest}(\bar \alpha)]$, where velocity oscillations can lead to arresting blocks, but where propagation will continue even if blocks behind the front arrest. The arresting region is determined by $\bar \tau (\bar \alpha) \in [0, \text{min} \left \{ \bar \tau_\text{arrest}(\bar \alpha),\bar \tau_\text{uncond}(\bar \alpha) \right \}]$. 

\subsection{Heterogeneous interfaces}
\begin{figure}
\centering
\includegraphics{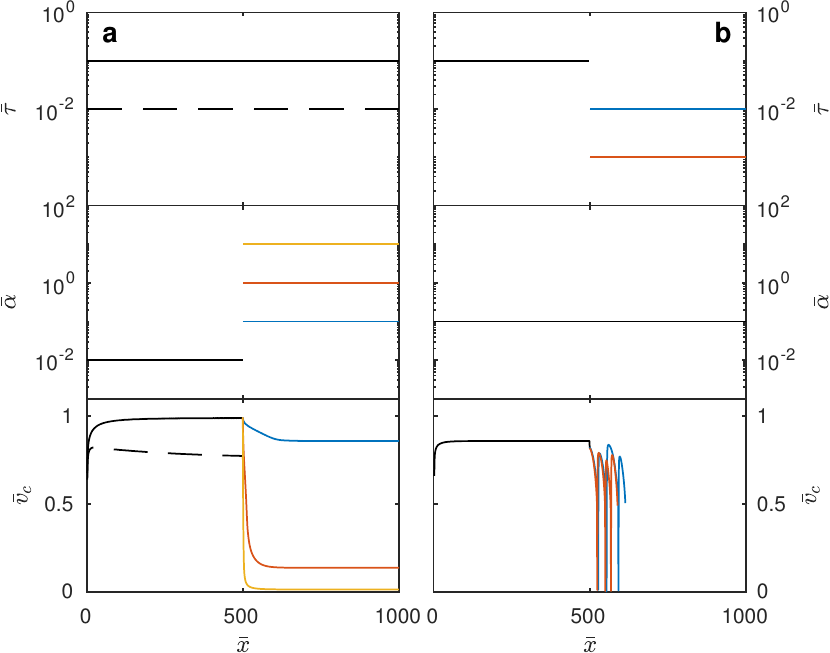}
\caption{Spatial transitions in $\bar \alpha$ and $\bar \tau$ can act as barriers to propagation. a: Step change in $\bar \alpha$ for a front that starts out as a slip pulse (dashed) and a crack (solid). If a slip pulse enters a region where slip pulses are not possible as steady state solutions, the front will arrest (abruptly within a few blocks, which is why the dashed line is discontinued at $\bar x \approx 500$), even though both regimes in $\bar \alpha$ allow for steady state front propagation. If the front starts out as a crack, propagation continues as long as the region it enters allows for steady state propagation. b: Step change in $\bar \tau$ so that the front enters the arresting region of the phase diagram. In this case, the front arrests through a series of start-stop events.
\label{fig:varyingTauAlpha}}
\end{figure}

The front type phase diagram of \figurename~\ref{fig:phasediagram}a has been constructed from steady-state data. Here we investigate to what extent it can be used to understand some features of fronts propagating along heterogeneous interfaces. \figurename\ref{fig:varyingTauAlpha} illustrates that transitions in $\bar \alpha$ and $\bar \tau$ can act as barriers to propagation, which can be understood from \figurename~ \ref{fig:phasediagram}a. 

Changes in $\bar \alpha$ can lead to arresting fronts if a front initiated in a region of $(\bar \tau, \bar \alpha)$ corresponding to steady state propagation enters a region corresponding to the arresting regime. This is demonstrated in the \figurename~\ref{fig:varyingTauAlpha}b, where fast cracks entering regions of smaller $\bar \tau$ arrest. In such cases, the criterion for start-stop fronts in equation \ref{eq:startstop} may be satisfied in the arrest phase, leading to multiple start-stop events before the motion stops completely. This is visible as velocity fluctuations after $\bar x = 500$ in the bottom row of \figurename~\ref{fig:varyingTauAlpha}b. As shown in \figurename~\ref{fig:varyingTauAlpha}a (dashed lines), if a front is initiated in the slip-pulse regime and then enters a region of larger $\bar \alpha$ crossing $\bar \tau_\text{uncond}$ (equation \ref{eq:uncond_prop}), it will arrest even if the region of larger $\bar \alpha$ corresponds to slow rupture. In the simulations in \figurename~\ref{fig:varyingTauAlpha}a, this arrest occurs within a few blocks. This means that a propagating front entering a region of different $\bar \alpha$ can arrest even though each value of $\bar \alpha$ would allow for a steady state propagation on a homogeneous interface.  For larger $\bar \tau$ where the entire interface is sliding when the region of increased $\bar \alpha$ is reached (\figurename~\ref{fig:varyingTauAlpha}a, solid lines), the front speed converges to a new value corresponding to the values of $\bar \tau$ and $\bar \alpha$ in that region of the phase diagram. 

\section{Discussion}
In this paper, we have demonstrated that a minimal one-dimensional model of rupture along frictional interfaces, obeying Amontons-Coulomb friction with velocity-strengthening dynamic friction, contains the main front types observed in the physics and geophysics literature.  This includes cracks, slip pulses and arresting fronts with steady state propagation speeds ranging from slow, to sub-Rayleigh and super-shear velocities. In addition to these steady state velocities, we observe fronts that alternate between arresting and propagating phases. This model can be written in terms of only two non-dimensional parameters that determine the front type. 
Complexity and richness of frictional rupture has been demonstrated to depend on different parameter ranges, boundary conditions, as well as spatial heterogeneities in stress constitutive parameters \cite{boatwright1996frictional,bizzarri2001solving,liu2005aseismic,bruhat2016rupture,helmstetter2009afterslip}. We emphasize that the observed complexity and richness of frictional rupture in this study occurs on interfaces that are homogeneous in both frictional properties and loading. This highlights that the large variation in modeled front types are likely generic features of frictional interfaces with velocity strengthening dynamic friction.

An important question is whether those results are robust against qualitative changes in the model; in particular whether there are specific effects related to the discontinuity of the friction law at vanishing slip speed. We have performed two additional sets of simulations using regularized friction laws; one with a velocity-weakening and one with a slip-weakening regularization. The corresponding equations of motion and the associated steady-state front type phase diagrams are presented in appendix \ref{sec:app_regularization}. Comparison between the phase diagram in \figurename~\ref{fig:phasediagram}a and the regularized models in \figurename~\ref{fig:phasediagram_regularized} indicates that most qualitative features are essentially unchanged. While some differences may be noted (details in appendix \ref{sec:app_regularization}), the spatial organization of the various regions in the phase diagram remains largely unchanged, showing that the discontinuity of the friction law does not change our main conclusions. 

We have also performed a set of two-dimensional simulations to address whether our results would be specific to the 1D nature of the system. We combine the bulk model of \cite{tromborg2011transition} with the present friction law. The parameters used and the results obtained are presented in Appendix \ref{sec:app_2D}. The obtained phase diagram (\figurename~\ref{fig:phasediagram2D}) is again similar to \figurename~\ref{fig:phasediagram}a, which demonstrates that our main conclusions are not artefacts of the 1D nature of the model.

The relative locations of the various regions in \figurename~\ref{fig:phasediagram} are consistent with experimental observations. At low $\bar \alpha$ and $\bar \tau$, the model predicts the existence of slip pulse solutions, in agreement with the experimental observation that slip pulses occur when the prestress is low compared to the static friction threshold \cite{nielsen2010experimental}. For $\bar \alpha = 0$, the model only predicts super-shear rupture \cite{amundsen2015steady}. For non-zero $\bar \alpha$, sub-Rayleigh and slow rupture can occur. Super-shear rupture can still occur if the prestress is large. Overall, the propagation speed increases with increasing prestress, which is consistent with experimental observations \cite{ben2010dynamics}.  Slow fronts have also previously been reported to depend strongly on velocity strengthening friction \cite{bar2013instabilities,bar2015velocity}. Here, slow propagation occurs at large $\bar \alpha$. Both the slip speed and the slow propagation speed are directly controlled by the velocity strengthening term $\bar \alpha$, leading to a slow propagation speed inversely proportional to $\bar \alpha$.

In addition to steady state rupture, the model predicts unsteady rupture velocities, where a crack alternates periodically between sub-Rayleigh speed and a transient arrest. Restarting arrested cracks requires that sufficient slow slip occurs in the broken part of the interface. Intermittent rupture then continues as long as the slow slip endures. A similar mechanism was found to control the transition from fast to slow rupture in a multi-asperity model \cite{tromborg2014slow}, reproducing observations in laboratory experiments \cite{rubinstein2004detachment}. We also speculate that the start-stop regime found in this study may be an analog to observed periodic pulsing of aseismic events have been observed \cite{nadeau2004periodic}.

In real systems, the prestress $\bar \tau$ can vary largely depending on the boundary conditions.  For side driven systems, the stress at the interface after a rupture has passed is likely to coincide with the dynamic friction level \cite{amundsen20121d,tromborg2011transition}, which corresponds to $\bar \tau \approx 0$.  This assumption is consistent with the observation in continuum rate-and-state models that the velocity corresponding to the minimum friction force sets the steady state slip speed and thus the rupture velocity \cite{bouchbinder2011slow}. In our simulations, this minimum is located at zero velocity. However, the possibility of a prestress that can be larger than the dynamic level leads to a large variety of possible rupture speeds. 

Several mechanisms can be responsible for varying stress conditions on frictional interfaces. Romanet et al. \cite{romanet2018fast} showed that the interaction between two fault planes can lead to the co-existence of sub-Rayleigh and slow rupture on the same fault. Interactions between fault planes could lead to large variations in the stress conditions of the fault planes prior to rupture. This is consistent with our findings for large $\bar \alpha$, where variations in $\bar \tau$ alone can lead to propagation speeds ranging from slow, through sub-Rayleigh to super-shear. 

Heterogeneities of the interface can also be due to spatial variations in the stress conditions or frictional properties. For instance, viscous patches along frictional interfaces have been shown to act as barriers to propagation because they can inhibit fast slip \cite{nielsen2010experimental}. Similarly, in our simulations, changes in $\bar \alpha$ and $\bar \tau$ along a frictional interface can cause rupture fronts to continue with a different velocity, or arrest, depending on whether the initial front propagates as a crack or a slip pulse, and on the region of the phase-diagram that the new value of $\bar \alpha$ and $\bar \tau$ corresponds to.

Our simulations show a region where rupture fronts will arrest, even when $\bar \tau > 0$. At low $\bar \tau$, this region causes a clear separation between sub-shear and slow rupture regions. In nature, observations show that fast and slow rupture obey different scaling relations between seismic moment and earthquake duration \cite{ide2007scaling}. There is currently an ongoing debate about whether there should exist a continuum of scalings between these two end-members \cite{ide2007scaling,peng2010integrated}. For prestress close to the dynamic threshold where $\bar \tau \approx 0$, the arresting region in $\bar \tau$ and $\bar \alpha$ could inhibit observations of intermediate rupture velocities, in turn causing observations of earthquake rupture mainly in the fast and slow end members.

\section{Conclusion}
In this paper, we have demonstrated that a minimal model of homogeneously loaded interfaces containing only two dimensionless parameters reproduces a wide range of observed slip and rupture behavior. This includes arresting fronts, slip pulses, unsteady rupture velocity, slow slip and rupture, fast rupture and super-shear rupture. Our results indicate that richness of frictional rupture is an inherent property of frictional systems with velocity strengthening dynamic branches.

\acknowledgments{K.T acknowledges support from EarthFlows - A strategic research initiative by The Faculty of Mathematics and Natural Sciences at the University of Oslo. H.A.S acknowledges support from the Research Council of Norway through the FRINATEK program, grant number 231621.}

\bibliographystyle{naturemag}
\bibliography{paper.bbl}

\appendix
\section{Equations of motion \label{sec:app_eom}}
The equation of motion for the one-dimensional Burridge-Knopoff model with a viscous term $\alpha \dot{u}_i$ is
\begin{equation}
    m\ddot{u}_i = k(u_{i+1}-u_i) + k(u_{i-1}-u_i) - \alpha_i \dot{u}_i - f_{f,i},
    \label{eq:EoM_1}
\end{equation}
where $i$ is the block index, $u$ is the displacement, $m$ is the mass, $k$ is the spring constant $\alpha$ is the viscosity coefficient, the blocks are separated by a distance $\Delta x$, and $f_f$ is the friction force. $f_f$ obeys Amontons-Coulomb law of friction, where a block $i$ begins to move when the static friction force $f_{f,\text{stuck},i} = \mu_{\text{s},i} p_i$ is reached. When moving, the friction force is $f_{f,\text{moving},i} = \mu_{k,i} p_i \dot u_i / |\dot u_i|$. A block arrests when $\dot u$ changes sign. Now assume that all blocks are initialized with positions $u_i(0)$. Any additional movement $u'_i(t)$ can be described by 
\begin{equation}
u_i(t) = u_i(0) + u'_i(t).
\label{eq:displacement}
\end{equation}
Combining equation \ref{eq:EoM_1} and \ref{eq:displacement} yields
\begin{equation}
    m\ddot{u}_i = k(u'_{i+1}-u'_i) + k(u'_{i-1}-u'_i) - \alpha_i \dot{u}'_i - f_{f,i} + \tau_i,
\end{equation}
where we have introduced the prestress
\begin{equation}
    \tau_i = k(u_{i+1}(0) - 2u_i(0) + u_{i-1}(0)).
\end{equation}
We then define the dimensionless variables $\bar{u} = \frac{u'}{U}$, $\bar{t}= \frac{t}{T}$ and $\bar x = \frac{x}{X}$ so that
\begin{equation}
    \ddot{\bar u}_i = \frac{kT^2}{m}(\bar u_{i+1}-2\bar u_i + \bar u_{i-1}) - \frac{\alpha_i T}{m} \dot{\bar u}_i - \frac{T^2}{mU}(f_{f,i} + \tau_i),
\end{equation}
where the derivative is now taken with respect to $\bar t$. Selecting
\begin{equation}
T = \sqrt{\frac{m}{k}}, \quad U = \frac{\mu_{\text{s},i} p_i- {\mu_{\text{k},i}} p_i}{k}, \quad X = \Delta x,
\end{equation}
we obtain
\begin{equation}
    \ddot{\bar{u}}_i -\bar{u}_{i-1} - \bar{u}_{i+1} + 2\bar{u}_i + \bar{\alpha}_i \dot{\bar{u}}_i - \bar\tau^\pm_i = 0
\end{equation}
where
\begin{equation}
\bar \alpha_i = \frac{\alpha_i}{\sqrt{km}}, \quad \bar \tau^\pm_i  = \frac{\tau_i/p_i \mp \mu_{\text{k},i}}{\mu_{\text{s},i} - \mu_{\text{k},i}},
\end{equation}
where $\pm$ corresponds to $\text{sign}(\dot {\bar u}_i)$. 

For most of this paper, we consider homogeneous interfaces ($\bar \alpha_i = \bar \alpha$, $\mu_{\text{k},i}=\mu_k$, $\mu_{\text{s},i}=\mu_s$) and homogeneous prestress ($\tau_i=\tau$, $p_i=p$). We also assume that the propagation is in the positive direction. This means that we set $\bar \alpha_i = \bar \alpha$ and $\bar \tau^+_i = \bar \tau$, obtaining equation \ref{eq:EoM}. Note that means that equation \ref{eq:EoM} is only valid for both positive and negative velocities in the special case when $\mu_\text{k}  = 0$. A small portion of the simulations we perform will contain oscillations with negative velocities (far) behind the front tip, and these results are thus only strictly valid under the assumption $\mu_\text{k} =0$. We have checked that this choice does not affect the propagation speed, but the detailed dynamics behind the front could depend on $\mu_\text{k}$. 
The constraint $p\mu_\text{s} \geq \tau$ results in the existence of steady state propagation only when $\bar \tau \in [0,1]$. The choice of $X$ also ensures that a dimensionless front propagation speed of $1$ corresponds to the velocity of sound in the system
\begin{equation}
\bar v_s = \Delta x \sqrt{\frac{k}{m}}  \frac{T}{X} = 1.
\end{equation}

Next, we set the boundary conditions. Block $1$ ruptures when the friction force reaches the static friction threshold. If the system is driven by a spring with spring constant $K$ driven at velocity $v$, this corresponds to adding a force on block 1, which in dimensionless units becomes $\bar F_\text{driving} =  1 - \bar \tau  +\bar K\bar v \bar t$, where $\bar K = \frac{K p}{\mu_\text{s}- \mu_k}$ and $\bar t = 0$ is the time when the first block reaches the static friction threshold. For soft tangential loading $\frac{\bar K\bar v}{\bar t} \ll 1$, this boundary condition is reduced to $\bar u_0 = 1-\bar \tau$.

In the Burridge-Knopoff model, the elastic modulus is given by $E = \frac{k\Delta x}{S}$, where $S$ is the cross-sectional area of the blocks. The mass density is defined as $\rho = \frac{m}{\Delta x S}$, which we make use of in the main text.

\section{Criterion for the unconditional existence of steady state propagation \label{sec:app_tau_uncond}}
If a block at the front tip is able to trigger the next block even though the block behind it has stopped, a propagating front will not be able to arrest. This criterion can be formulated as follows:
The minimum criterion in $\bar\tau(\bar\alpha)$ for the existence of a steady state propagation is that a block stops at exactly $\bar{u} = (1-\bar{\tau})$, corresponding to the static friction threshold of the next block, thus triggering it.
This assumptions translates to $\bar{u}_{i-1} = 1-\bar{\tau}$, $\dot {\bar{u}}_{i-1} = 0$, $\bar{u}_{i+1} = 0$, $\dot {\bar{u}}_{i+1} = 0$. From equation \ref{eq:EoM} we find
\begin{equation}
    \ddot {\bar{u}}(\bar{t}) + \bar{\alpha} \dot{\bar{u}} + 2\bar{u} - 1 = 0,
\end{equation}
which has the solution
\begin{equation}
    \bar{u}(\bar t) = c_1 e^{\frac{1}{2} \left ( -\sqrt{\bar{\alpha}^2-8} - \bar{\alpha} \right )\bar t} + c_2 e^{\frac{1}{2} \left (\sqrt{\bar{\alpha}^2-8} - \bar{\alpha} \right )\bar t} + \frac{1}{2}.
\end{equation}
From the assumptions $\dot{\bar{u}} (0) = 0$ and $\bar u (0) = 0$ we find
\begin{align}
    \bar{u}(\bar{t}) & =  [ \frac{1}{2\sqrt{8-\bar \alpha^2}}\sin \left ( \frac{\sqrt{8-\bar \alpha^2}}{2}\bar t \right ) \\
&    - \frac{1}{2}\cos \left ( \frac{\sqrt{8-\bar \alpha^2}}{2}\bar t \right ) ] e^{-\frac{\bar \alpha}{2}\bar t} + \frac{1}{2},\nonumber
\end{align}
where we have assumed that the system is underdamped ($\bar \alpha \leq 2 \sqrt{2}$). 
From $\bar u (\bar t) = 1- \bar\tau$ we have
\begin{align}
\label{eq:x_t}
& \frac{1}{\sqrt{8-\bar \alpha^2}}\sin \left ( \frac{\sqrt{8-\bar \alpha^2}}{2}\bar t_s \right ) - \cos \left ( \frac{\sqrt{8-\bar \alpha^2}}{2}\bar t_s \right ) \\
& = (1-2\bar \tau )e^{\frac{\bar \alpha}{2}\bar t_s} \nonumber
\end{align}
where $\bar t_s$ is the time at which the block position reaches $1-\bar \tau$. The requirement of zero velocity at $\bar t= \bar t_s$ can be found from $\dot {\bar u} (\bar t = \bar t_s) = 0$
\begin{equation}
\frac{1}{\sqrt{8 - \bar\alpha^2}} 2\sin \left ( \frac{\sqrt{8 -\bar  \alpha^2}}{2} \bar t_s \right ) e^{-\frac{\alpha}{2} \bar t_s} = 0,
\end{equation}
where we are looking for the first non-trivial solution
\begin{equation}
\bar t_s = \frac{2 \pi}{\sqrt{8-\bar \alpha^2}}.
\end{equation}
Inserting for $\bar t_s$ in equation \ref{eq:x_t} we obtain
\begin{equation}
\bar \tau = \frac{1}{2} -\frac{1}{2}e^{-\frac{\pi \bar \alpha}{\sqrt{8 - \bar \alpha^2}}}
\end{equation}
which gives us the line of unconditional propagation in the phase diagram of $\bar \tau$ and $\bar \alpha$, as shown in \figurename~\ref{fig:phasediagram}.

\section{Two-dimensional simulations \label{sec:app_2D}}
Here we address whether our results would be specific to the 1D nature of the system.  We performed a set of simulations in 2D with the spring-block model described in \cite{tromborg2011transition}. We simulate a slider of dimensions $(L,H) = (0.9,0.015)$ m, with $(600 \times 10)$ blocks. We use friction coefficients $\mu_\text{s} = 0.4$ and $\mu_\text{k} = 0.2$, and a varying velocity strengthening term $\alpha$. We use a Young's modulus $E = 3$ GPa, density $\rho = 1300$ kgm$^{-3}$, width $w = 0.006$ m, with a bulk damping coefficient of $\nu = \sqrt{0.1 k m}$. To limit wave reflections from the top surface, we use a damping term $\nu = \sqrt{km}$ at the top blocks. The normal force on the bottom blocks is prescribed to $1$ kN per block, and the system is initialized with a prestress $\tau = \tau_\text{init} + (1-\tau_\text{init})e^{-3x/H}$, and the slider is pushed from all blocks on the left interface. The system is solved using adaptive time-stepping and event detection for the transition from static to dynamic friction. The simulations are run until all blocks have ruptured or all blocks have arrested. 

\begin{figure}
\centering
\includegraphics{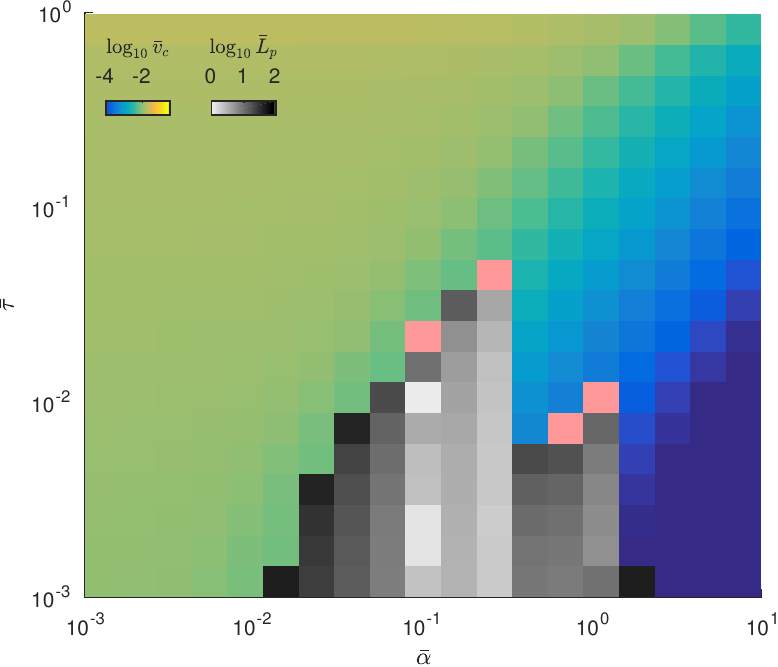}
\caption{Front velocities from two-dimensional simulations. $\bar v_c$ is given as the measured front velocity (measured as the average front velocity between $L/2$ and $L$) over the shear wave speed of the material. $\bar L_p$ (greyscale) is the propagation length $L_p$ over the system height $H$. Start-stop fronts are marked with pink.
\label{fig:phasediagram2D}}
\end{figure}

\figurename~\ref{fig:phasediagram2D} shows the resulting front velocities, which confirm that the qualitative behavior from the one-dimensional still remains in two dimensions, and that the main conclusions are not artefacts of the one-dimensional nature of the model.'

\section{Regularization of the friction law \label{sec:app_regularization}}
An important question is whether the results from the main text are robust against qualitative changes in the model. In particular, one may first ask whether there is any specific effect related to the discontinuity of the friction law at vanishing slip speed, when the frictional resistance on a block abruptly drops from the static friction force to the dynamic friction force. To answer the question, we performed two additional sets of simulations, using regularized friction laws: one with a velocity-weakening and one with a slip-weakening regularization. 

\begin{figure*}
\centering
\includegraphics{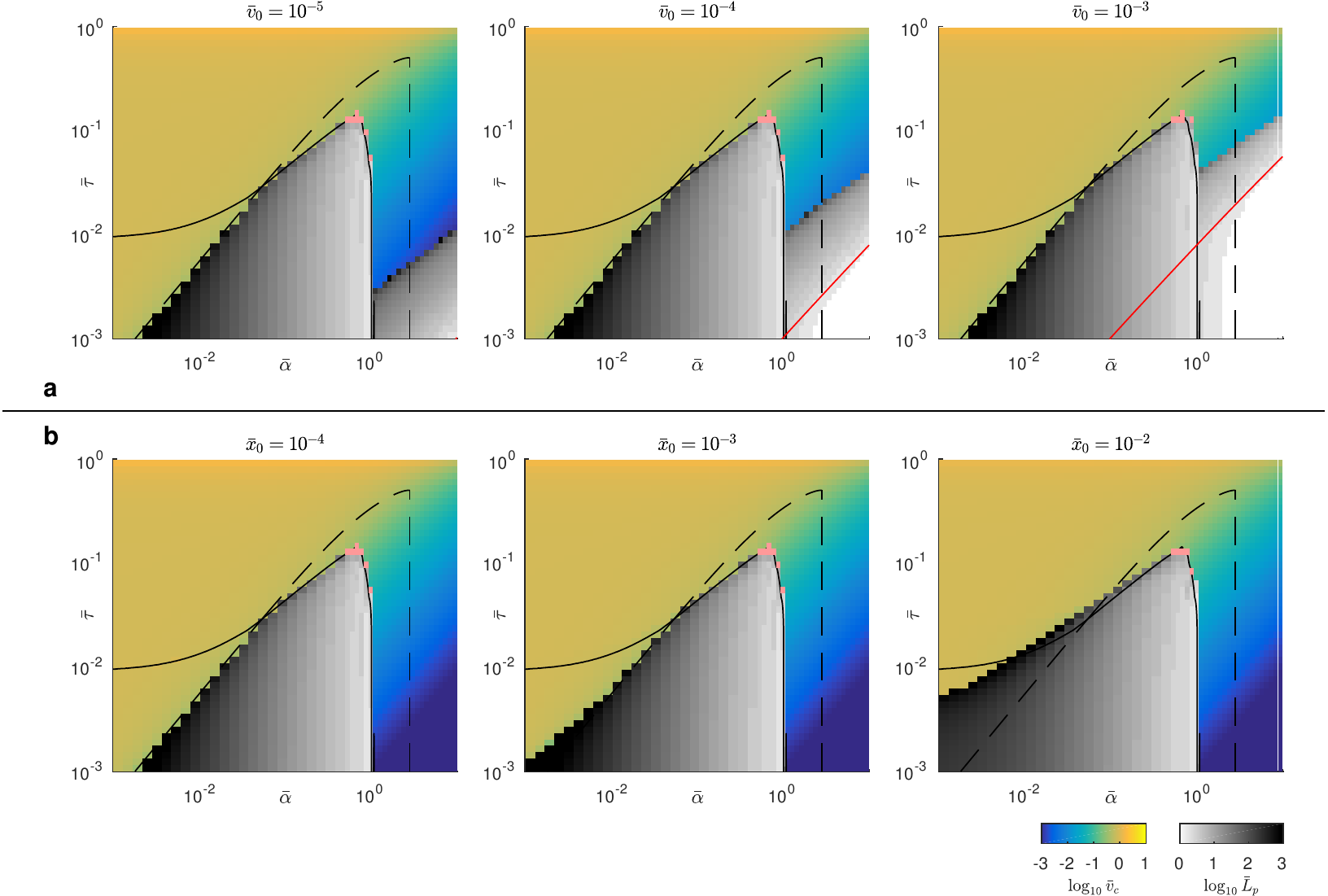}
\caption{Front velocities using regularized friction with (a) velocity regularization from equation \ref{eq:EoMRegaularizedVelocity} (the red line corresponds to equation \ref{eq:regularized_phasediagram_from_minimum}) and (b) displacement regularization from equation \ref{eq:EoMRegaularizedDisplacement}. Start-stop fronts are marked with pink. The grayscale colormap show the propagation length $L_p$ of arresting fronts. The solid and dashed lines are the same as in \figurename~\ref{fig:phasediagram}.
\label{fig:phasediagram_regularized}}
\end{figure*}
First, we introduce a velocity scale for the decay from static to dynamic friction so that the equation of motion for sliding blocks can be written as 
\begin{align}
    \label{eq:EoMRegaularizedVelocity}
    \ddot{\bar{u}}_i -\bar{u}_{i-1} - \bar{u}_{i+1} + 2\bar{u}_i + \bar{\alpha} \dot{\bar{u}}_i + \frac{\dot {\bar u}_i}{|\dot {\bar u}_i|}e^{-\frac{|\dot {\bar{u}}_i|}{\bar{v}_0}}- \bar\tau \\
    = 0, \quad \forall i \in [1,N] \nonumber
\end{align}
where $\bar v_0$ is a characteristic velocity scale that we vary. The resulting front velocities and propagation lengths are shown in \figurename~\ref{fig:phasediagram_regularized}a. The main effect of the velocity regularization is that it introduces a minimum in the friction law that gives a minimum $\bar \tau_\text{min} (\bar \alpha)$ that allows for steady state propagation. For this criterion, which is the main cause of arresting in the large $\bar \alpha$ regime, we can immediately set up the expression
\begin{equation}
\bar \tau_\text{min} (\bar \alpha) = \bar v_0 \bar \alpha (1 - \log (\bar v_0 \bar \alpha)).
\label{eq:regularized_phasediagram_from_minimum}
\end{equation}
This line is shown in red in \figurename~\ref{fig:phasediagram_regularized}. No steady state can exist below this curve.

We also perform regularization with a displacement dependent term, which results in the following equation of motion:
\begin{align}
    \label{eq:EoMRegaularizedDisplacement}
    \ddot{\bar{u}}_i -\bar{u}_{i-1} - \bar{u}_{i+1} + 2\bar{u}_i + \bar{\alpha} \dot{\bar{u}}_i + \frac{\dot {\bar u}_i}{|\dot {\bar u}_i|}e^{-\frac{|\dot {{\Delta u}}_i|}{\bar{x}_0}}- \bar\tau \\
    = 0, \quad \forall i \in [1,N], \nonumber
\end{align}
where $\Delta u_i$ is the displacement of block $i$ since the last time it ruptured.   As can be seen in \figurename~\ref{fig:phasediagram_regularized}b, the phase diagram is essentially unaffected by the regularization for small $\bar x_0$. For larger $\bar x_0$, the slip weakening regularization induces a widening of the arresting region at small $\bar \alpha$ values when the characteristic slip distance large. In those cases, the main effect is to shrink the region where slip pulses are allowed, making them more difficult to identify as a potential front type in the model.

Comparison between the phase diagram of the main model (\figurename~\ref{fig:phasediagram}a) and that of the regularized models (\figurename~\ref{fig:phasediagram_regularized}) indicates that most qualitative features are essentially unchanged. In particular, the spatial organisation of the various regions (front types) in the phase diagrams are unchanged, showing that the discontinuity of the friction law does not change our main conclusions.

\end{document}